%% file: sample-sigconf.tex
\documentclass[sigconf]{acmart}
\usepackage{nopageno}
\usepackage{savefnmark}

\usepackage{booktabs} 

\usepackage[T1]{fontenc}
\usepackage{graphicx}
\usepackage{amssymb}
\usepackage{graphicx}
\usepackage{tabularx}
\usepackage{multicol}
\usepackage{multirow}
\usepackage{booktabs}
\usepackage{threeparttable}
\usepackage{adjustbox}
\usepackage{epstopdf}
\usepackage{amsmath}
\usepackage{mathrsfs}
\usepackage{algorithm}
\usepackage{algorithmicx}
\usepackage{algpseudocode}
\usepackage{subfig}
\usepackage{url}
\usepackage{color}
\usepackage{amsmath,amsfonts,amssymb}
\usepackage{amsmath}
\DeclareMathOperator*{\argmax}{arg\,max}
\setcopyright{acmcopyright}

\newcolumntype{P}[1]{>{\centering\arraybackslash}p{#1}}

\copyrightyear{2019} 
\acmYear{2019} 
\acmConference[CIKM '19]{The 28th ACM International Conference on Information and Knowledge Management}{November 3--7, 2019}{Beijing, China}
\acmBooktitle{The 28th ACM International Conference on Information and Knowledge Management (CIKM '19), November 3--7, 2019, Beijing, China}
\acmPrice{15.00}
\acmDOI{10.1145/3357384.3357903}
\acmISBN{978-1-4503-6976-3/19/11}

\begin{document}
	
	\fancyhead{}
	
	\title{Multi-Hot Compact Network Embedding}

	\author{Chaozhuo Li} 
	\affiliation{Beihang University}
	\email{lichaozhuo@buaa.edu.cn}
	
	\author{Lei Zheng}
	\authornote{Both authors contributed equally to the paper}
	\affiliation{University of Illinois at Chicago}
	\email{lzheng21@uic.edu}
	
	\author{Senzhang Wang}
	\affiliation{Nanjing University of Aeronautics and Astronautics}
	\email{szwang@nuaa.edu.cn}
	
	\author{Feiran Huang}
	\affiliation{Jinan University}
	\email{huangfr@jnu.edu.cn}
	
	\author{Philip S. Yu} 
	\affiliation{University of Illinois at Chicago}
	\email{psyu@uic.edu}
	
	\author{Zhoujun Li}
	\affiliation{Beihang University}
	\email{lizj@buaa.edu.cn}

	\begin{abstract}
	Network embedding, as a promising way of the network
	representation learning, is capable of supporting various subsequent network mining and analysis tasks, and has attracted growing research interests recently.
	Traditional approaches assign each node with an independent continuous vector, which will cause  memory overhead for large networks. 
	In this paper we propose a novel multi-hot compact network embedding framework to effectively reduce memory cost by learning partially shared embeddings.
	The insight is that  a node embedding vector is composed of  several basis vectors according to a multi-hot index vector.
	The basis vectors are shared by different nodes, which can significantly reduce the number of continuous vectors while maintain similar data representation ability.
	Specifically, we propose a MCNE$_{p}$ model to learn compact embeddings from pre-learned  node features. A novel component named compressor is integrated into MCNE$_{p}$ to tackle the challenge that popular back-propagation optimization cannot propagate loss through discrete samples.
	We further propose an end-to-end model MCNE$_{t}$ to learn compact embeddings from the input network directly.
	Empirically, we evaluate the proposed models over four real network datasets, and the results demonstrate that our proposals can save about 90\% of memory cost of network embeddings without significantly performance decline. 
	\end{abstract}

	\maketitle
	
	\input{samplebody-conf}
	
	\bibliographystyle{ACM-Reference-Format}
	\bibliography{sample-bibliography}
	
\end{document}

%% file: samplebody-conf.tex
\section{Introduction}

Nowadays, information networks are becoming ubiquitous in the real-world life, including social networks (e.g., Twitter, LinkedIn), paper citation networks (e.g., DBLP, Arxiv) and knowledge graphs (e.g., Freebase, Wikipedia).
Mining valuable knowledge from these information networks is crucial to a variety of real applications in practice. 
A fundamental problem in the area of graph mining is learning a desirable representation vector for each node \cite{cao2015grarep}, which is called network embedding. 
The learned high-quality node representations are critically important  to perform many down-streaming data mining tasks, such as node classification \cite{grover2016node2vec,cui2018survey}, social recommendation \cite{hu2016clustering,zhang2018network} and link prediction \cite{perozzi2014deepwalk}.

\begin{figure}[!tp]
	\centering
	\subfloat[One-hot index.]{
		\centering
		\label{fig:framework_tr} 
		\includegraphics[width=26.5mm]{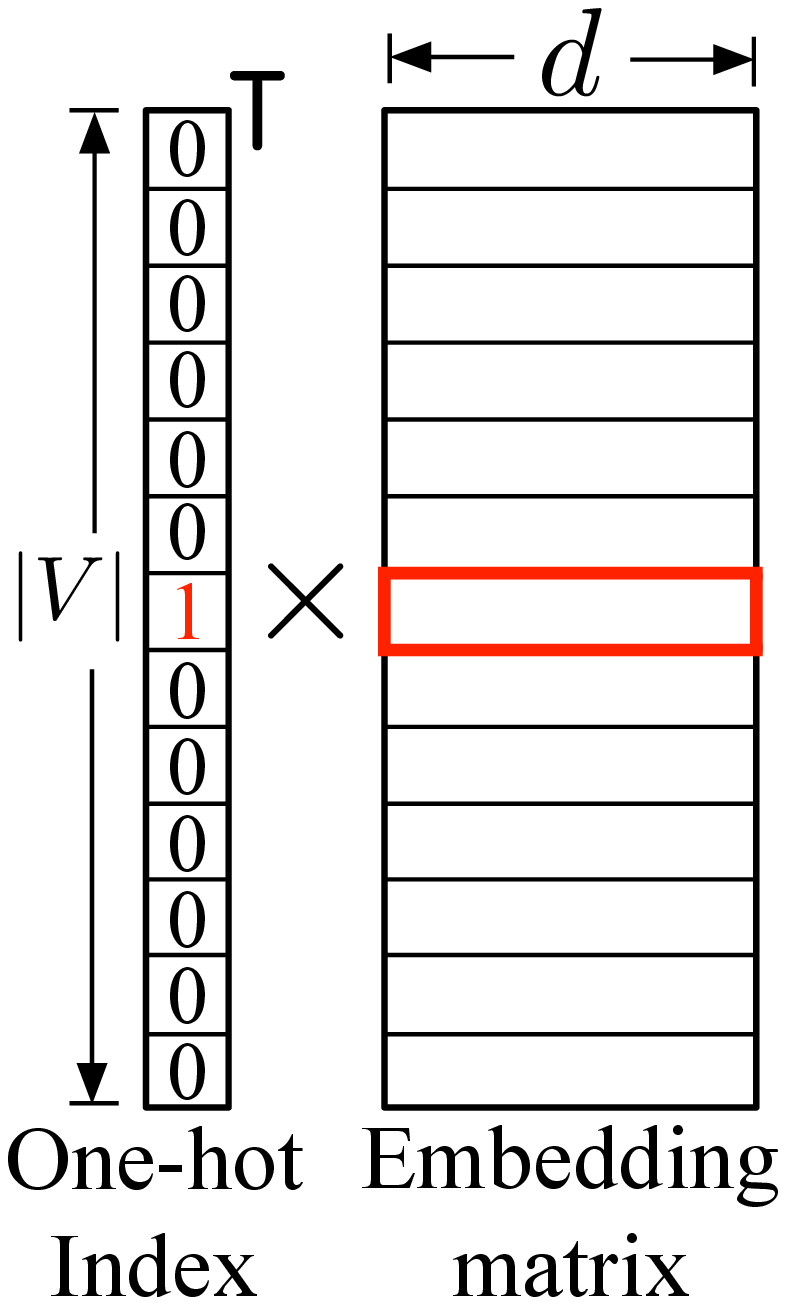}}
	\hspace{0.001in}
	\subfloat[KD coding.]{
		\centering
		\label{fig:framework_kd} 
		\includegraphics[width=26.5mm]{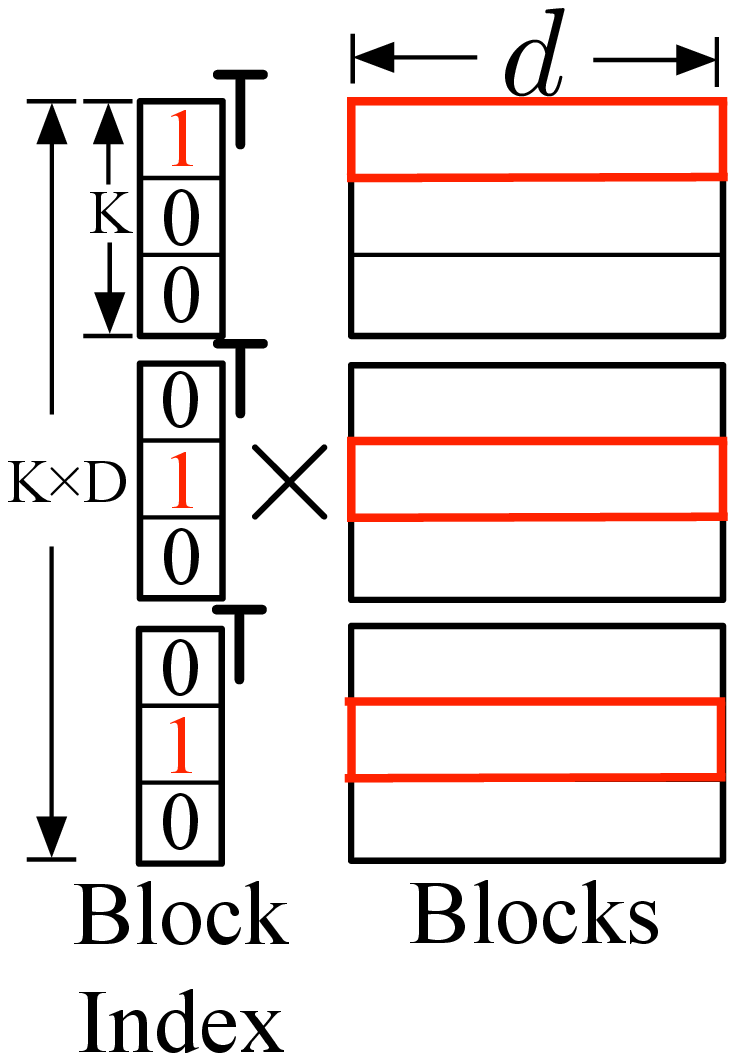}}
	\hspace{0.001in}
	\subfloat[Multi-hot index.]{
		\centering
		\label{fig:framework_ml} 
		\includegraphics[width=26.5mm]{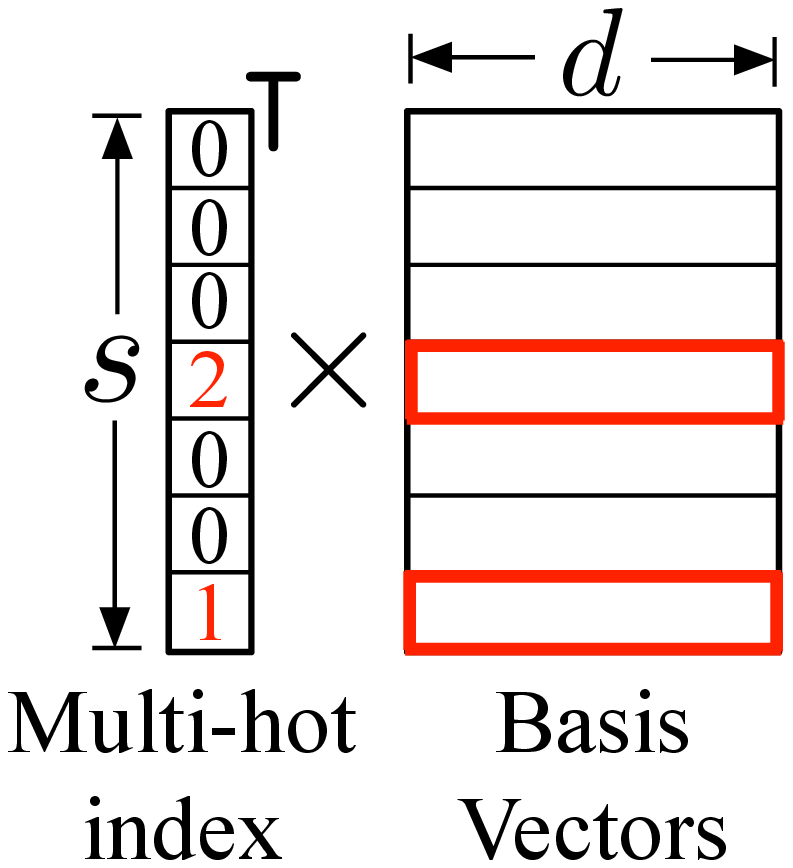}}
	\caption{Comparison between three embedding strategies.}
	\label{fig:aes} 
\end{figure}

Most existing network embedding methods \cite{perozzi2014deepwalk,tang2015line,grover2016node2vec,wang2016structural,yang2015network} encapsulate the input network into a continuous embedding matrix.
Each row in the embedding matrix is the learned representation of a node.
However, as each node is assigned an independent embedding vector, the size of the embedding matrix can be huge for large networks. 
Given a network with $|V|$ nodes and the dimension $d$ of the node embeddings, the learned embedding matrix has ($|V| \times d$) real-valued entries.
When applied to downstream data mining tasks, the entire embedding matrix needs to be loaded into the memory as the lookup table, which will cause significant memory overhead. 
Even for a network of medium size (100,000 nodes)  and $d = 500$ (a popular choice), the  embedding matrix has 50 million float entries , which will cost more than 1 GB memory. 
The high storage cost will be a key bottleneck to many memory sensitive scenarios, such as an educational application in the mobile platform may need to load the knowledge graphs to provide accurate information.

In this paper we study the novel problem of learning compact network embeddings which require much less memories while do not sacrifice the performance too much.
The straightforward memory saving strategies include reducing the dimension $d$ or representing nodes with discrete vectors instead of the continuous ones such as DNE \cite{Shen2018Discrete}. However, both strategies will seriously reduce the information preserving ability of the embeddings, which leads to the significant performance sacrifice.
For example,  the learned embeddings from DNE are composed of binary codings instead of real-valued numbers, and thus DNE can reduce the memory cost. However, when DNE is performed in the unsupervised learning scenario, it only achieves half of the accuracy obtained by DeepWalk in node classification on Flickr and Youtube datasets \cite{Shen2018Discrete}.
This is mainly because the representation ability of the  binary codes is quite limited. 
For example in the 2-dimensional space, the  binary codes can only utilize the four points ((-1,1),(1,-1),(-1,-1),(1,1)) to present the nodes, while the continuous features can exploit the whole plane to present the data.


Traditional network embedding strategies can be understood as the multiplication between the one-hot index vector and the embedding matrix as shown in Figure \ref{fig:framework_tr}. 
The major limitation of this one-hot index method is that each node needs to be represented as a unique id and an independent embedding vector, which may cause redundancy. 
For example if two nodes are connected and share many common neighbors, their representations tend to be similar and assigning two independent vectors for them will cause resource waste. 
One-hot index method needs $|V|$ continuous vectors to represent the network with $|V|$ nodes.
Recently, Chen et al. \cite{chen2018learning} proposed KD coding to reduce the memory cost of embeddings. As shown in Figure \ref{fig:framework_kd}, KD coding consists of D blocks and each block contains K basis vectors.
For each node, KD coding firstly selects a basis vector from each block according to the block indexes, and then combines the selected D vectors as the final node representation.
By sharing the basis vectors, KD coding can compose $K^{D}$  different feature vectors.
A comparatively  small K and D can compose into a large number of (larger than $|V|$) representation vectors.
It shows that KD coding is able to significantly reduce the memory cost without much performance drop  \cite{chen2018learning}.
However, the introducing of blocks limits the compression capacity of KD coding because the blocks reduce the sharing ratios of the basis vectors. 
Each component of the final node representation can only be selected from K basis vectors in a specific block and has nothing to do with the rest K $\times$ (D-1) basis vectors in other blocks, which will cause resource waste.

To address the above issues of KD coding, in this paper we propose a multi-hot compact network embedding strategy to further improve the compression rate. Our idea is shown in Figure \ref{fig:framework_ml}. 
Each node is associated with a $s$-dimensional multi-hot index vector, where $s$ is the number of basis vectors and the sum of non-zero elements in the multi-hot index vector is $t$. 
The embedding vector of a node is composed by $t$ component vectors, which are selected from $s$ basis vectors according to the multi-hot index. 
By removing the blocks in KD coding, the proposed multi-hot index model enables each component vector of the final node representation can be selected from all the basis vectors instead of the partial ones, which improves the sharing ratios of basis vectors.   
Besides, each basis vector can be only selected at most once in KD coding, while our method allows duplicate selection.
Given the same number of basis vectors for KD coding and the multi-hot method ($s = K \times D$ and $t = D$), the representation space of our method is $s^{t}$, while the KD coding can only achieve $(\lfloor s/t \rfloor)^{t}$.

The studied multi-hot network embedding problem is difficult to address due to the following two challenges.
First, the multi-hot indexes of the nodes are trainable discrete parameters, which are difficult to be learned by existing network embedding methods. 
For the matrix factorization based models, it is difficult to formally define the multi-hot indexes as the discrete subjections to guide the direction of matrix factorization process \cite{chen2018learning}, while for the neural network based models, it is intractable to calculate the derivations of the discrete indexes in the back-propagation optimization process \cite{jang2016categorical}.
Second, as a number of network embedding models have been proposed, it is time and effort efficient to directly transform the one-hot embeddings pre-learned by existing models  to the multi-hot compact embeddings.
A desirable multi-hot embedding  framework should be able to handle both the traditional end-to-end  scenario and the  mentioned pre-learned feature compressing scenario.

In this paper, we propose a Multi-hot Compact Network Embedding (MCNE) framework to reduce the memory cost of the network embeddings. 
MCNE is a deep auto-encoder based model consisting of three components: the encoder, the compressor and the decoder.
Encoder transforms the inputs to the latent vectors in the pre-defined dimension, and then feeds them into the compressor.
Then the compressor generates the discrete multi-hot indexes from the latent vectors, which introduces the gumbel-softmax trick to back propagate the loss through the discrete neural nodes.
Finally, decoder generates the final node representations according to the learned multi-hot indexes and the basis vectors.
Specifically, we first propose the MCNE$_{p}$ model to compress pre-learned one-hot index based features into the multi-hot embeddings. Considering the inputs of MCNE$_{p}$ are the formatted distributed vectors, its encoder is implemented by a Multi-Layer Perceptron (MLP) to ensure the model efficiency. 
Then we also propose MCNE$_t$ model for the end-to-end learning scenario. 
The network topology is viewed as the model input under this scenario, which is more sophisticated (e.g., sparse and high-dimensional). 
In order to capture the highly non-linear topology information, we introduce the graph convolutional network (GCN) as the encoder of MCNE$_{t}$.  
The compressor and decoder of MCNE$_{t}$ are same to the ones in MCNE$_{p}$.
We conduct extensive experiments over four datasets on two tasks.
Experimental results demonstrate that MCNE$_{p}$ can save about 90\% of memory cost on average without significant performance decline, and MCNE$_{t}$ outperforms the baselines with less memory cost. 

We summarize our main contributions as follows.
\begin{itemize}
	\item We propose a novel and flexible multi-hot network embedding framework MCNE, which can be  applied on both scenarios of the pre-learned feature compressing and the end-to-end learning from scratch with little modifications.
	
	\item We design a novel  compressor to learn discrete multi-hot indexes for the nodes, which tackles the challenge that current back-propagation optimization methods cannot propagate loss through discrete samples.
	
	\item  Extensively, we evaluate the proposed models on four datasets. The results show the superior performance of our proposals by the  comparison with state-of-the-art baseline methods.
\end{itemize}
The rest of this paper is organized as follows. 
Section 2 gives a brief review on related works. 
Then we formally define the studied problem in section 3.
In section 4 we introduce the details of proposed MCNE$_{p}$ model.
The end-to-end embedding model MCNE$_{t}$ is presented in section 5. 
Experimental results are shown and discussed in section 6. Finally, we give concluding remarks in section 7.

\section{Related Work}\label{RelatedWork}


Network embedding is closely related to the manifold learning problem \cite{roweis2000nonlinear, tenenbaum2000global, belkin2003laplacian,li2006very}.
Manifold learning aims to perform dimensionality reduction on the high-dimensional datasets. 
Traditional manifold learning methods usually convert the input dataset into a feature matrix, and then apply the matrix factorization techniques on the feature matrix to obtain the eigenvectors. 
Manifold learning models are usually designed for general usage, which may ignore the unique characteristics in the network topology.

Most existing network embedding models focus on encapsulating the topology information into the node representations \cite{zhang2018cosine,pan2018adversarially,ma2018hierarchical,kim2018side,chen2018harp}. The motivation is that nodes with similar topology structures (e.g., many common neighbors) should be distributed closely in the learned latent space. 
DeepWalk \cite{perozzi2014deepwalk} converted the network into a set of node sequences by random walk, and then the skip-gram model was applied on the node sequences to generate node embeddings.
As an improved version of DeepWalk, Node2Vec \cite{grover2016node2vec} proposed a more flexible loss function to capture the second-order proximity. 
Ribeiro et al. proposed struc2vec \cite{ribeiro2017struc2vec} to generate the embeddings of node structural identities.
SDNE \cite{wang2016structural} was a deep auto-encoder based model, which exploited the fine tuning strategy to capture the global structural information.
Recently several works introduced the generative adversarial network (GAN) into network embedding learning \cite{dai2017adversarial,wang2017graphgan} to stimulate the underlying true connectivity distribution over all the nodes.

Besides the network topology information, several works focused on incorporating the side information as the complementary to improve the quality of node embeddings \cite{tu2016max,yang2015network,li2017Semi,rossi2018deep,liu2018content,kim2018side}. 
Tu et al. \cite{tu2016max} extended the DeepWalk model to a max-margin extension to incorporate few available labels.
Li et al. \cite{li2017Semi} designed a multi-layer perceptron based model to perform semi-supervised network embedding.
Yang et al. proposed TADW \cite{yang2015network}, a matrix factorization based model to fuse node attributes (text features) into the embedding process. 
DANE \cite{gao2018deep} manually calculated the  correlations between  structural information and the node attributes, and then integrated them into an unified objective.

Although a lot of network embedding models have been proposed, they usually suffer from high memory usage. Recently some works have focused on learning memory saving embeddings \cite{chen2018learning,Shen2018Discrete,zhang2018cosine} . 
DNE \cite{Shen2018Discrete} learned binary codings as the node embeddings by adding binary subjections into the matrix factorization, which suffered from undesirable embedding quality. 
KD \cite{chen2018learning}  was the first work to learn compact embeddings from the pre-learned features, whose performance was limited by the block strategy.
Besides, existing models are designed for a specific scenario (pre-learned feature compressing \cite{chen2018learning} or end-to-end learning \cite{Shen2018Discrete, zhang2018cosine}), but there still lacks a more general framework that can handle both cases. 

\section{Problem Definition}
A network is denoted as $G = \{V, T\}$, in which $V$ is the node set, and $T \in \mathbb{R}_{\{0,1\}}^{|V| \times |V|}$ is the adjacency matrix. 
We formally define the studied problem as follows:
\begin{definition}
	\textbf{Multi-hot Compact Network Embedding.} 
	Given the input network $G$ and the representation dimension $d$, we aim to learn a multi-hot index matrix $H \in \mathbb{N}^{|V| \times s}$ and a basis matrix $B \in \mathbb{R}^{s \times d}$, where $s$ is the number of basis vectors.
	Each row in $B$ is a shared basis vector.
	Each row $H_{i} \in \mathbb{N}^{1 \times s}$ in matrix $H$ represents the multi-hot index vector of the node $i$, which is under the restriction of $\sum_{j=1}^{s}H_{ij} = t$. 
	The final embedding matrix $E \in \mathbb{R}^{|V| \times d}$ is composed by the multiplication between matrix $H$ and $B$. 
	
\end{definition}
The restriction of matrix $H$ ensures the embedding vector is composed by $t$ selected basis vectors (duplicate selection is allowed).  	
The learned embedding matrix $E$ should satisfy that, in the learned latent space,  the feature vectors of nodes with similar topologies would be distributed close to each other.

\section{Multi-hot  Network Embedding}

\subsection{Compact Network Embedding from Pre-learned Features}
\label{pre}
We first present the MCNE$_{p}$ model which learns the multi-hot compact embeddings from the pre-learned features.
We assume that the node embeddings have been learned by traditional models and applied on the industry applications.
Considering that it is time and resource consuming to re-learn the embeddings of the entire network from scratch,  we aim to compress the pre-learned one-hot index based embeddings with MCNE$_{p}$ model.
Besides, there exists various types of networks (e.g., the signed networks and attributed networks). 
By converting the pre-learned embeddings into the compact ones, MCNE$_{p}$ can be applied on different types of networks without needing to know their unique characteristics, which proves the generality of our proposal.

\begin{figure}
	\centering
	\includegraphics[width=0.4\textwidth]{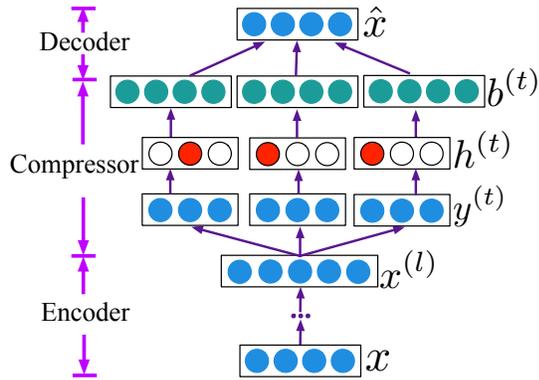}
	\caption{An illustration of the proposed MCNE$_{p}$ model.}
	\label{fig:mcne} 
\end{figure}

Given the learned one-hot index based embedding matrix $E_{p} \in \mathbb{R}^{|V| \times d}$, we aim to learn the multi-hot index matrix $H$ and the shared basis matrix $B$ under the constraint that the multiplication of $H$ and $B$ is similar to $E$ as much as possible.  
This task can be viewed as the reconstruction of the original embeddings, where the auto-encoder model thrives on \cite{Li2018SSDMV}. Hence we propose MCNE$_{p}$, an auto-encoder based deep model, to convert the original embeddings into the compact ones.
As shown in Figure \ref{fig:mcne}, the proposed MCNE$_{p}$ model has three major components: encoder, compressor and decoder. 
Given an original embedding vector $x \in \mathbb{R}^{1 \times d}$, MCNE$_{p}$ will proceed  the following steps.  

\noindent \textbf{Encoder} 
Encoder transforms the input vector $x$ to a latent vector $x^{(l)} \in \mathbb{R}^{1 \times d_{l}}$, in which $l$ is the index of the top layer and $d_{l}$ is the number of neural cells in layer $l$. 
Formally, given the input $x$, we perform the following calculations in the encoder:
\begin{align}
x^{(1)} &= \phi(W^{(1)}\cdot x + b^{(1)})        \\
x^{(k)} &= \phi(W^{(k)}\cdot x^{(k-1)} + b^{(k)}), k=2,\cdots,l
\end{align}
where $W^{(k)}$ is the weight matrix, $b^{(k)}$ is the bias vector and $x^{(k)}$ is the learned latent vector in the $k$-th layer. 
$\phi$ is the activation function which introduces the nonlinearity into the feature learning process. 
Here we select $tanh$ as the activation function because it has stronger gradients and can avoid the bias in the gradients \cite{lecun2012efficient}. 

\noindent \textbf{Compressor}
Compressor learns the multi-hot index vector $h$ for the input $x$, and then selects the basis vectors according to the learned index vector.
The sum of elements in the learned multi-hot index vector is $t$, which means the input $x$ can be reconstructed by $t$ basis vectors selected from all $s$ candidates according to the multi-hot indexes. 
Instead of directly sampling the multi-hot index vector, we sample $t$ one-hot vectors $\{h^{(1)},h^{(2)},\dots,h^{(t)}\}$ and then combine them together. The two sampling strategies are equivalent when the duplicate sampling is allowed. 

\begin{figure}
	\centering
	\includegraphics[width=0.35\textwidth]{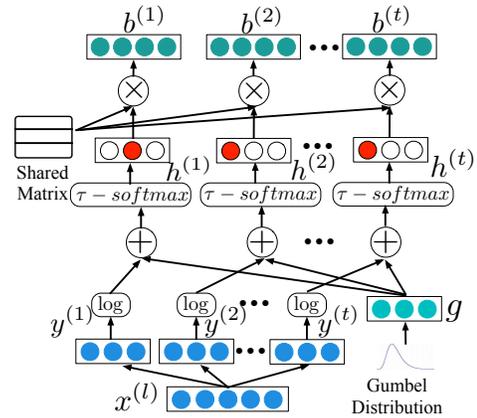}
	\caption{An illustration of the compressor.}
	\label{fig:compressor} 
\end{figure}

As shown in the bottom of Figure \ref{fig:compressor}, given the output $x^{(l)}$ from the encoder, compressor first converts it into $t$ vectors $y^{(i)} \in \mathbb{R}^{1 \times  s}, 1 \leqslant i \leqslant t$. This process contains two steps: first linearly transform the $x^{(l)}$ into a $s \times t$ vector, and then reshape the learned vector to a $ \mathbb{R}^{s \times t}$ matrix. The $i$-th row of the learned matrix is  $y^{(i)}$:
\begin{align}
[y^{(1)},y^{(2)}, \dots, y^{(t)}]^\intercal = \phi(\gamma(W^{(c)} \cdot x^{(l)} + b^{(c)}))
\end{align}
in which $\gamma$ is the reshape operation. We select $softplus$ function as the  activation function $\phi$ to ensure the elements in  $y^{(i)}$ are positive.

Each $y^{(i)} \in \mathbb{R}^{1 \times  s}$ is viewed as an independent categorical distribution, which has $s$ categories and the probability of the category $k$ is proportional to $y^{(i)}_{k}$: $\Pr(h^{(i)}_{k} = 1) \propto y^{(i)}_{k}$. 
The sampled category $k$ represents that the $k$-th basis vector is selected. 
In this way, each node has its unique $t$ categorical distributions.
Note that the linear transformation and the activation functions ($tanh$ and $softplus$) are all monotone functions, which ensure the vectors $y^{(i)}$ of similar nodes are also similar. Hence, similar nodes will share closer categorical distributions,  leading to the similar index vectors.

We aim to sample a discrete parameter from each categorical distribution $y^{(i)}$ to form up the multi-hot index. For example, if the category 1 is sampled from $y^{(0)}$ and the category 4 is sampled from $y^{(1)}$, the multi-hot index will be (1, 4) and the input will be constructed by the 1-$st$ and 4-$th$ basis vectors.
To tackle the challenge that back-propagation optimization cannot propagate loss through discrete samples, we introduce the gumbel-softmax trick \cite{maddison2016concrete}, which is a popular re-parameterization technique to generate discrete samples by adding a stochastic noise from the standard gumbel distribution.

%
%
%

Gumbel-softmax first samples a noise vector $g \in \mathbb{R}^{1 \times s}$  from a  standard gumbel distribution, and then add it with $y^{(k)}$ as $z = \log y^{(k)} + g$.
We can prove the $\argmax_{i}  (z_{i})$ operation is equivalent to sampling a discrete sample from the categorical distribution $y^{(k)}$ as follows.
Firstly we calculate the marginal probability when $z_{m}$ is the largest entry in $z$:
\begin{equation}
\begin{aligned}
\Pr(\max_{i \neq m} z_{i} < z_{m}) =& \prod_{i \neq m} \Pr((\log y^{(k)}_{i} + g_{i}) < (\log y^{(k)}_{m} + g_{m}))  \noindent \\
=& \prod_{i \neq m} \Pr(g_{i} < (\log y^{(k)}_{m} + g^{t} - \log y^{(k)}_{i})) \\
=& \prod_{i \neq m} e^{-e^{-(z_{m} - \log y^{(k)}_{i}))}} \nonumber
\end{aligned}
\end{equation}
Based on the marginal distribution over $z_{m}$, we need to integrate it out to find the overall probability:
\begin{equation}
\begin{aligned}
& \Pr(m = \argmax_{i}(z_{i})) \\
&= \int e^{-(z_{m} -\log y^{(k)} _{m} )-e^{-(z_{m} -\log y^{(k)}_{m})}} \prod_{i \neq m} e^{-e^{-(z_{m} - \log y^{(k)}_{i}))}} \mathrm{d} z_{m} \\
& =\frac{y^{(k)}_{m}}{\sum^{s}_{i=1}y^{(k)}_{i}} \nonumber
\end{aligned}
\end{equation}
Hence, the distribution of the maximum element's position in the vector $z$ is equivalent  to the  normalized categorical distribution $y^{(k)}$.
However, $\argmax$ operator is not differentiable, so the $\tau$-softmax function  \cite{jang2016categorical} is introduced as a continuous approximation of $\argmax$ operator:
\begin{equation}
h^{(k)}_i = \frac{\text{exp}((\log(y^{(k)}_i)+g^{i})/\tau)}{\sum_{j=1}^s \text{exp}((\log(y^{(k)}_j)+g^{j})/\tau)},  \quad i=1, ..., s \nonumber
\end{equation}
where $\tau$  is the temperature parameter to control how closely the $\tau$-softmax function approximates the $\argmax$ operation. As $\tau \to 0$, the softmax becomes an $\argmax$ and $h^{(k)}$ is sampled from the categorical distribution $y^{(k)}$. During training, we first let $\tau > 0$ to allow gradients past the neural cells, and then gradually reduce the temperature $\tau$.

As shown in the bottom right corner of Figure \ref{fig:compressor}, a noise vector $g$ is sampled from a standard gumbel distribution. The gumbel distribution can be achieved from a uniform distribution by inverse transform sampling \cite{vogel2002computational}:
\begin{equation}
g \sim -log(-log(\text{Uniform}(0,1))).
\end{equation}
Then $t$ approximate one-hot vectors $h^{(i)} \in \mathbb{R}^{1 \times s}$ can be achieved by the following calculation process:
\begin{equation}
h^{(i)} = \tau-softmax(\log(y^{(i)}) + g),  \quad i=1, ..., t.
\end{equation}
The positions of maximum entries in $h^{(i)}$ can be combined as the final multi-hot indexes.
After that, $t$ basis vectors will be selected from $s$ candidates according to the sampled one-hot indexes:
\begin{equation}
b^{(i)} = h^{(i)} \times B
\end{equation}
The shared matrix $B$ contains $s$ basis vectors with the dimension of $d$. The entries of shared matrix $B$ are trainable parameters which are randomly initialized. 
The basis vectors will be updated in the model training step.

\noindent \textbf{Decoder} 
Given the  the selected basis vectors $b^{(i)}$ from  compressor, decoder composes them into the reconstructed vector $\hat{x}$. 
We choose the plus operation to combine the selected basis vectors:
\begin{equation}
\label{decoder}
\hat{x} = \sum_{i=1}^{t} b^{(i)}.
\end{equation}
When the learned compacting embeddings are utilized in downstream applications such as node classification, the selected basis vectors are also need to be combined as the final node representation in the same way. Hence we select the simple but fast plus operator to ensure the efficiency.

\noindent \textbf{Objective Function} 
The objective of the MCNE$_{p}$ is to make the composed vector $\hat{x}$ similar to the original embedding $x$ as much as possible. We choose to use the following reconstruction loss as the objective function:
\begin{equation}
\label{formual:1}
\mathcal{L} = \frac{1}{|V|} \sum_{i=1}^{|V|} \lVert x_{i} - \hat{x}_{i} \rVert ^{2}
\end{equation}

\noindent \textbf{Model Export}  
After the model training finished, we need to export the basis matrix and the multi-hot index vector of each node. 
The matrix $B$ in the compressor is saved as the basis matrix. 
For each node, we can get its corresponding one-hot vectors $h^{(1)}, h^{(2)}, \dots, h^{(t)}$ from the compressor. 
For each vector $h^{(i)}$, we get the position of its maximum element as a single number: $ \argmax_{k} h^{(i)}_{k}$. 
After processing  all the  one-hot vectors, we can get $t$ integers in the range of $[1,s]$ as the multi-hot indexes of the input node.
With the learned multi-hot indexes and shared basis vectors, the compact node embeddings can be easily generated following the Formula \ref{decoder}.

\subsection{End-to-end Compact Network Embedding}
In this subsection we introduce the MCNE$_{t}$ model to learn the multi-hot compact network embeddings from the scratch.
Given a network G with its adjacency matrix $T \in \mathbb{R}^{|V| \times |V|} $,   MCNE$_{t}$  aims to learn the multi-hot embeddings from the network topology directly. 
The input topology is more sophisticated than the pre-learned features exploited in the MCNE$_{p}$ model, and thus MCNE$_{t}$ should own stronger feature learning ability. 
Recently graph convolutional network (GCN) is popular as it nicely integrates local
node features and graph topology in the convolutional layers  \cite{kipf2016semi}.
GCN follows a neighborhood aggregation scheme, where the representation vector of a node is computed by recursively aggregating and transforming representation vectors of its neighboring nodes \cite{kipf2016semi}.
The original GCN is designed for the semi-supervised node classification task, while we further extend it to perform the unsupervised network embedding task.
MCNE$_{t}$ learns the multi-hot indexes based on the convoluted features from GCN, and utilizes the connectivity information as the indicator to guide the learning of final compact embeddings.
Similar to MCNE$_p$, From bottom to top MCNE$_{t}$ includes three major components: GCN, compressor and decoder.

\noindent \textbf{GCN} 
GCN performs convolution operation on the topology structure, which is essentially a first-order approximation of localized spectral filters on the networks \cite{wu2019comprehensive}.
Given the adjacency matrix $T$, GCN performs the following operation in the $k$-th layer:
\begin{align}
G^{(k+1)} &= \phi(\tilde{D}^{-\frac{1}{2}}\tilde{T}\tilde{D}^{-\frac{1}{2}}G^{(k)}W^{(k)}).      
\end{align}
$\tilde{T} = T + I_{|V|}$ is the adjacency matrix with added self-connections, in which 
$I_{|V|}$ is the identity matrix. $\tilde{D}_{ii}=\sum_{j}\tilde{T}_{ij}$ and other elements in $\tilde{D}$ are zeros. $W^{(k)}$ is a layer-specific trainable weight matrix.
$G^{(k)} \in \mathbb{R}^{|V| \times d_{k}}$ is the input feature matrix of $k$-th layer and $d_{k}$ is the pre-defined dimension.
Usually the node attributes are viewed as the initial input matrix $G^{(0)}$.
As we focus on the topology-based network embedding task, the input matrix $G^{(0)}$ is randomly initialized and its entries are viewed as the trainable parameters which can be updated in the model learning process. 
The proposed MCNE$_{t}$ can be easily applied on the attributed networks by setting matrix $G^{(0)}$ as the node attributes.
$tanh$ is selected as the activation function $\phi$. 
The output vector $g_{i}$ from the top layer of GCN is viewed as the latent vector of node $n_{i}$.

\noindent \textbf{Compressor} From the top layer $l$ of GCN we can obtain a latent feature matrix $G = G^{(l)} \in \mathbb{R}^{|V| \times d_{l}}$. 
Each row $g_{i}$ in $G$ is the latent vector of the corresponding node $n_{i}$.
The compressor is similar to the one in MCNE$_{p}$ model but only differs in the inputs.
The compressor in MCNE$_{p}$ considers the outputs from the encoder as the inputs, while the one in MCNE$_{t}$ views the learned latent features $G$ from GCN as the inputs.
From the compressor, we can obtain the multi-hot indexes of the input nodes.

\noindent \textbf{Decoder} Same to the one  of MCNE$_{p}$, decoder selects the basis vectors according to the multi-hot indexes and adds them as the reconstructed vector.
Given the input node $n_{i}$, decoder generates the reconstructed version $\hat{g}_{i}$ of the latent vector $g_{i}$. 
The outputs from decoder are viewed as the final compact embeddings.

\noindent \textbf{Objective function} The loss of MCNE$_{t}$ includes two parts: a reconstruction loss $\mathcal{L}_{r}$ to ensure the reconstructed vector $\hat{g}_{i}$ should be similar to its original version $g_{i}$, and a topology-guided loss $\mathcal{L}_{t}$ to ensure the learned compact embeddings can capture the structural connectivity information. 

The reconstruction loss is similar to the Formula \ref{formual:1}. 
The learned compact embedding $\hat{g}_{i}$ is expected to be similar to the latent feature $g_{i}$ learned from GCN, and thus the final compact embeddings can better capture the localized topology information.
This loss is formally defined as:
\begin{equation}
\mathcal{L}_{r} = \frac{1}{|V|} \sum_{i=1}^{|V|} \lVert g_{i} - \hat{g}_{i} \rVert ^{2}
\end{equation}

In addition, the learned compact embeddings are also expected to preserve the connectivity structural information. 
Hence we design another loss function $\mathcal{L}_{t}$ to capture the connectivity information, which also contributes to learning the trainable parameters in the GCN part. 
Given an input node $n_{i}$, we randomly select a neighbor node of $n_{i}$ as the positive sample $pos_i$, and an unconnected node as the negative sample $neg_i$. 
We further propose a pairwise negative sampling based objective function $\mathcal{L}_{t}$  to maximize the difference between the two node pairs: $<n_{i}, pos_{i}>$ and $<n_{i}, neg_{i}>$. 
The similarity between the connected nodes should be larger than the similarity between unconnected nodes as much as possible. $\mathcal{L}_{t}$ is formally defined as:
\begin{align}
\label{re}
\mathcal{L}_{t} = \frac{1}{|V|} \sum_{i=1}^{|V|} - \ln (\hat{g_{i}}^\intercal \hat{g}_{pos_{i}} - \hat{g_{i}}^\intercal \hat{g}_{neg_{i}})
\end{align}
By minimizing the loss $\mathcal{L}_{t}$, we can maximize the difference between the connected nodes and unconnected ones, which captures the connectivity structural information.
The final objective function of MCNE$_{t}$ is the weighted combination of these two losses:
\begin{align}
\label{mcnet}
\mathcal{L} = \mathcal{L}_{t} + \beta \cdot  \mathcal{L}_{r}
\end{align}
in which $\beta$ is the hyper-parameter to control the weight of reconstruction loss.
After training the model, we can obtain the multi-hot indexes of nodes and the shared basis vectors from the compressor, which can be utilized to construct the final compact embeddings.

\section{Experiments}
In this section, we will evaluate the performance of the proposed models on four real network datasets through two tasks.

\begin{table}
	\centering
	\small
	\begin{threeparttable}
		\caption{Statistics of the three datasets (Avg$_{t}$ denotes the average
			number of links per node).}
		\begin{tabular}{ccccc}
			\toprule
			\multicolumn{1}{c}{Dataset}&\multicolumn{1}{c}{Nodes} & \multicolumn{1}{c}{Edges}  & \multicolumn{1}{c}{Avg$_{t}$}&
			\multicolumn{1}{c}{Categories}\cr
			\midrule
			BlogCatalog &10,312 & 333,983 & 32.39 & 39 \cr
			DBLP &16,753 & 10,4371 & 6.23 & 8 \cr
			Flickr & 23,664 & 1,734,334 & 73.29 & 10 \cr
			Youtube & 1,138,499 & 2,990,443 & 2.63 & 47 \cr
			
			\bottomrule
		\end{tabular}
		\label{tab:statistics_of_datasets}
	\end{threeparttable}
\end{table}

\noindent \textbf{Datasets}
We utilize four datasets to evaluate the performance of different methods.
Table \ref{tab:statistics_of_datasets} shows the statistics of  four datasets. 

\begin{itemize}
	\item \textbf{BlogCatalog}\footnote{http://socialcomputing.asu.edu/pages/datasets} \saveFN\sft\ 
	In the BlogCatalog platform, users can post blogs and follow other users with similar interests.
	Nodes in this network are the users, and the edges are the following relationships between users.
	Users can join different interest groups, which are viewed as their categories.
	
	\item \textbf{DBLP}\footnote{https://aminer.org/citation} DBLP is an academic citation network. The nodes in the DBLP network are the manuscripts, and the edges are the citation relationships between papers.
	This dataset contains 16,753 papers from 8 research areas, which are viewed as the labels of the nodes.

	\item \textbf{Flickr}\useFN\sft\
	Flickr is an image sharing platform, in which users can post images and share them with friends. 
	The following relationships between users form the network structure.
	Users also can join different interest groups, which are considered as the node categories.
	
	\item \textbf{Youtube}\useFN\sft\
YouTube is a popular video sharing website.
Users can upload videos and follow other users with similar interests, which forms a social network. In this dataset,
nodes are users and labels represent the interest groups
	
\end{itemize}

\begin{table}
	\centering
	\small
	\begin{threeparttable}
		\caption{Parameter settings of KD ($K$ and $D$) and MCNE ($s$ and $t$).}
		\begin{tabular}{c|c|c|c|c}
			\toprule
			 Dataset & $K$ & $D$ & $s$ & $t$ \cr
			\midrule
			BlogCatalog &16 & 8 & 128 & 8 \cr
			DBLP & 16 & 8 & 128 & 8 \cr
			Flickr & 16 & 16 & 256 & 16 \cr
			Youtube & 256 & 32 & 8192 & 32 \cr
			\bottomrule
		\end{tabular}
		\label{tab:settings}
	\end{threeparttable}
\end{table}

\noindent \textbf{Baseline Methods}
To thoroughly evaluate the performance of our proposals, we select the following two types of baselines.
We compare the proposed models with the first type of baselines as follows to evaluate the compression ability:
\begin{itemize}
	\item \textbf{DNE} \cite{Shen2018Discrete} is an end-to-end model which learns binary coding based embeddings by adding binary subjections into the matrix factorization process.
	
	\item \textbf{KD} \cite{chen2018learning}  is the first work to learn compact embeddings from the pre-learned features, which is the strongest baseline for compressing the embeddings.
\end{itemize}
We also adopt the second type of baselines as follows to evaluate the quality of embeddings learned by the end-to-end MCNE$_{t}$ model:
\begin{itemize}
	\item \textbf{DeepWalk} \cite{perozzi2014deepwalk} first converts the network topology into a set of node sequences by random walk, and then the skip-gram model is applied on the generated node sequences to learn the node embeddings.
	
	\item \textbf{LINE} \cite{tang2015line} designs two loss functions to capture the first-order and second-order proximities. We concatenate the embeddings learned by the two objectives as the node features.
	
	\item \textbf{Node2Vec} \cite{grover2016node2vec} is an extension of DeepWalk by adding the guided random walk to capture the connectivity
	patterns.
	
	\item \textbf{SDNE} \cite{yang2015network} is a deep auto-encoder based model to capture the highly non-linear topology information. It aims to preserve both the local and global structural similarities.
\end{itemize}

\noindent \textbf{Parameter Settings}
The dimension of the node embeddings is set to 256 (a popular choice). 
For DNE, the portion of the labeled nodes $\lambda$  is set to $0$ as we aim to evaluate its performance in an unsupervised manner. 
For DeepWalk, LINE and Node2Vec, we follow the parameter settings in the original papers. 
The hyper-parameter $\alpha$ in SDNE is fixed to 0.5. The number of layers and the count of neural cells in each layer of SDNE are the same as the MCNE$_{p}$ model. 

To conduct a fair comparison, we set the numbers of basis vectors in MCNE and KD the same, which should satisfy $s = K \times D$ and $t=D$.
The settings of core parameters in MCNE ($s$ and $t$) and KD ($K$ and $D$) are shown in Table \ref{tab:settings}.
For the MCNE$_{p}$ model, the number of layers in the encoder and decoder is set to 2 and the node count in the hidden layer is set to $s/2$. The learning rate is set to 0.001. 
The parameter $\tau$ in the compressor is initialized as 1 and will decrease by 0.1 after every 100 epochs. The minimum value of $\tau$ is set to 0.5. 
The reconstruction weight $\beta$ in Formula \ref{mcnet} is set to 0.3. The batch size is set to 128 and the number of epochs is set to 500. 
For MCNE$_{p}$ model, at the end of each epoch, we evaluate the loss on a small validation set. The parameters with the lowest validation reconstruction loss are saved.
Hyper-parameters are tuned using the random search strategy on the validation dataset.
For MCNE$_{t}$ model, the checkpoint model with the lowest training loss will be saved as the final model. 
The number of GCN layers in MCNE$_{t}$ is set to 2, and the number of cells in the hidden layer is set to 1000.

\noindent \textbf{Evaluation methods} 
To thoroughly evaluate the performance of different embedding methods, we select node classification and link prediction as the downstream tasks for testing.
For node classification, the learned embeddings are viewed as the node features, and a one-$vs$-rest logistic regression classifier is trained to predict the node labels. 
$T_{r}$ percentages of nodes are randomly selected as the training set and the remaining nodes are the test samples.
Following the previous work \cite{perozzi2014deepwalk}, we set  $T_{r} = 10\%$ for the BlogCatalog, DBLP and Flickr datasets, and $T_{r} = 1\%$ for the Youtube dataset.
The F1-scores ( Micro-F1 and Macro-F1) are selected as the  metrics.

For link prediction, 30\% of edges are randomly removed from the original network.
Network embedding methods are applied on the remained subnetwork to learn node embeddings.
Node pairs from the removed edges are viewed as the positive samples.
We randomly sample the same number of node pairs which are not connected as the negative samples.
The cosine similarity score between two nodes of a node pair is calculated based on the learned embeddings.
The Area Under Curve (AUC) score is  selected as the metric to evaluate the consistency between the annotations and the calculated similarity scores of the node pairs.
We train each embedding method on each dataset five times and report the average results.
The one-$vs$-rest classifier, F1-scores and AUC score are implemented by the   scikit-learn tool\footnote{https://scikit-learn.org/stable/}.

\subsection{Evaluation on Compact Embedding from Pre-learned Features} 
In this subsection we will evaluate the performance of our model in compact network embedding from pre-learned features. 
Here we only compare MCNE$_{p}$ with the KD coding method, because other baselines focus on the end-to-end learning and thus are not comparable in this case.
The pre-learned node features are generated by the state-of-the-art network embedding model Node2Vec. 
The learned compact embeddings should be similar to the original features as much as possible, which means a good compression model should achieve comparable performance to Node2Vec.

\noindent \textbf{Quantitative Analysis } 
Table \ref{tab:compress} presents the quantitative performance of KD and MCNE$_{p}$ methods, where the results of Node2Vec are also shown as the benchmarks. 
Besides, the numbers of parameters and the memory costs along with the corresponding compression ratios are also reported.

As shown in Figure \ref{fig:framework_tr}, node embeddings learned by Node2Vec consist of two parts.
The first part is a $|V| \times d$ continuous embedding matrix, where $V$ is the count of nodes and $d$ is the dimension of embeddings.
The second part is the one-hot indexes representing a node with an integer, and thus it contains $|V|$ integers.
Overall, embeddings learned from Node2vec have ($|V| \times d + |V| \time 1$) parameters. 
In the python 2.7 of Linux system, the size of a float object is 16 bytes and an integer object costs 12 bytes. Thus Node2vec needs ($|V| \times d \times 16 + |V| \times 12$) bytes to store the learned embeddings. 
The memory overhead increases linearly with the number of nodes, which brings severe challenges to the memory-sensitive applications.
For the Youtube dataset, it has 292.59 million parameters and costs nearly 4GB memories. 

\begin{table}
	\centering
	\small
	\begin{threeparttable}
		\caption{Performance comparison of KD and MCNE$_{p}$ given the same memory costs.}
		\begin{tabular}{P{1.5cm}|P{1.5cm}|P{1.5cm}|P{0.7cm}P{1.5cm}}
			\toprule
			&Dataset& Node2Vec & KD& MCNE$_{p}$ \cr
			\midrule
			Node  &Blog.&0.371&0.353&	\textbf{0.369}\cr
			Classi.&DBLP&0.302&0.273&\textbf{0.309}\cr
			(Micro-F1)&Flickr&0.358&0.331&	\textbf{0.354}\cr
			&Youtube&0.385 &0.352&	\textbf{0.382}\cr
			\midrule
			Node  &Blog.&0.224&0.192&	\textbf{0.219}\cr
			Classi. &DBLP&0.126&0.101&\textbf{0.124}\cr
			(Macro-F1)&Flickr&0.162&0.144&\textbf{0.161}\cr
			&Youtube&0.306&0.267&	\textbf{0.301}\cr
			\midrule
			Link &Blog.&0.781&0.752&	\textbf{0.784}\cr
			Prediction &DBLP&0.591&0.562&\textbf{0.587}\cr
			(AUC) &Flickr&0.683&0.667&\textbf{0.680}\cr
			&Youtube&0.475&0.447&\textbf{0.469}\cr
			\midrule
			\# of &Blog.&2.65&0.12&	0.12 (22.08)\cr
			Parameters &DBLP&4.30&0.17&0.17 (25.29)\cr
			(Million) &Flickr&6.08&0.44&0.44 (13.82)\cr
			&Youtube&292.59&38.53&	38.53 (7.59)\cr
			\midrule
			\# of  Memory&Blog.&40.40&1.44&1.44 (28.06)\cr
			(MB) &DBLP&65.63&2.03&2.03 (32.33)\cr
			&Flickr&92.71&5.33&5.33 (17.39)\cr
			&Youtube&4,460.29&448.93&	448.93 (9.94)\cr
			\bottomrule
		\end{tabular}
		\label{tab:compress}
	\end{threeparttable}
\end{table}

\begin{figure}[!tp]
	\centering
	\subfloat[Blogcatalog.]{
		\centering
		\label{fig:blog} 
		\includegraphics[width=41mm]{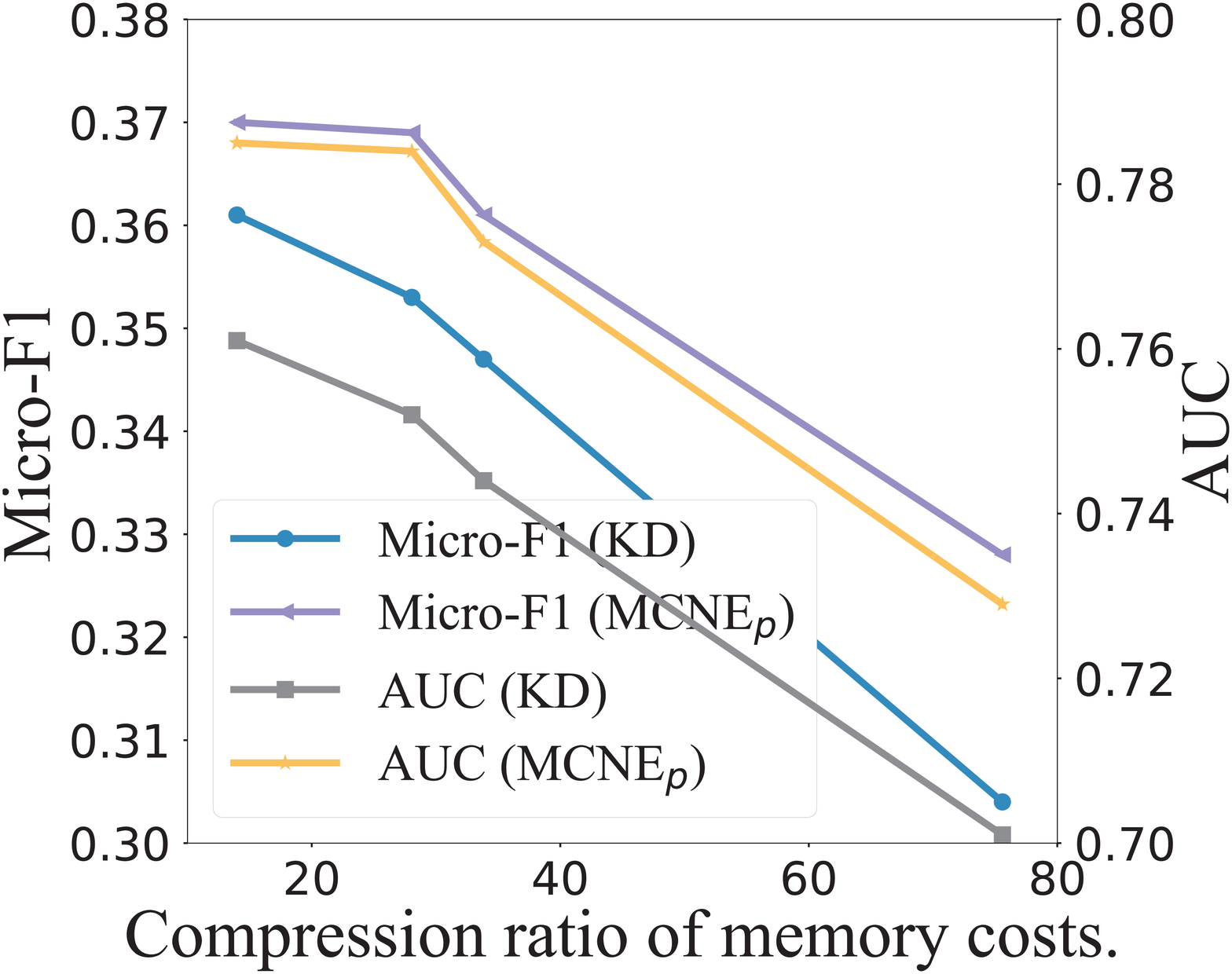}}
	\hspace{0.001in}
	\subfloat[DBLP.]{
		\centering
		\label{fig:dblp} 
		\includegraphics[width=41mm]{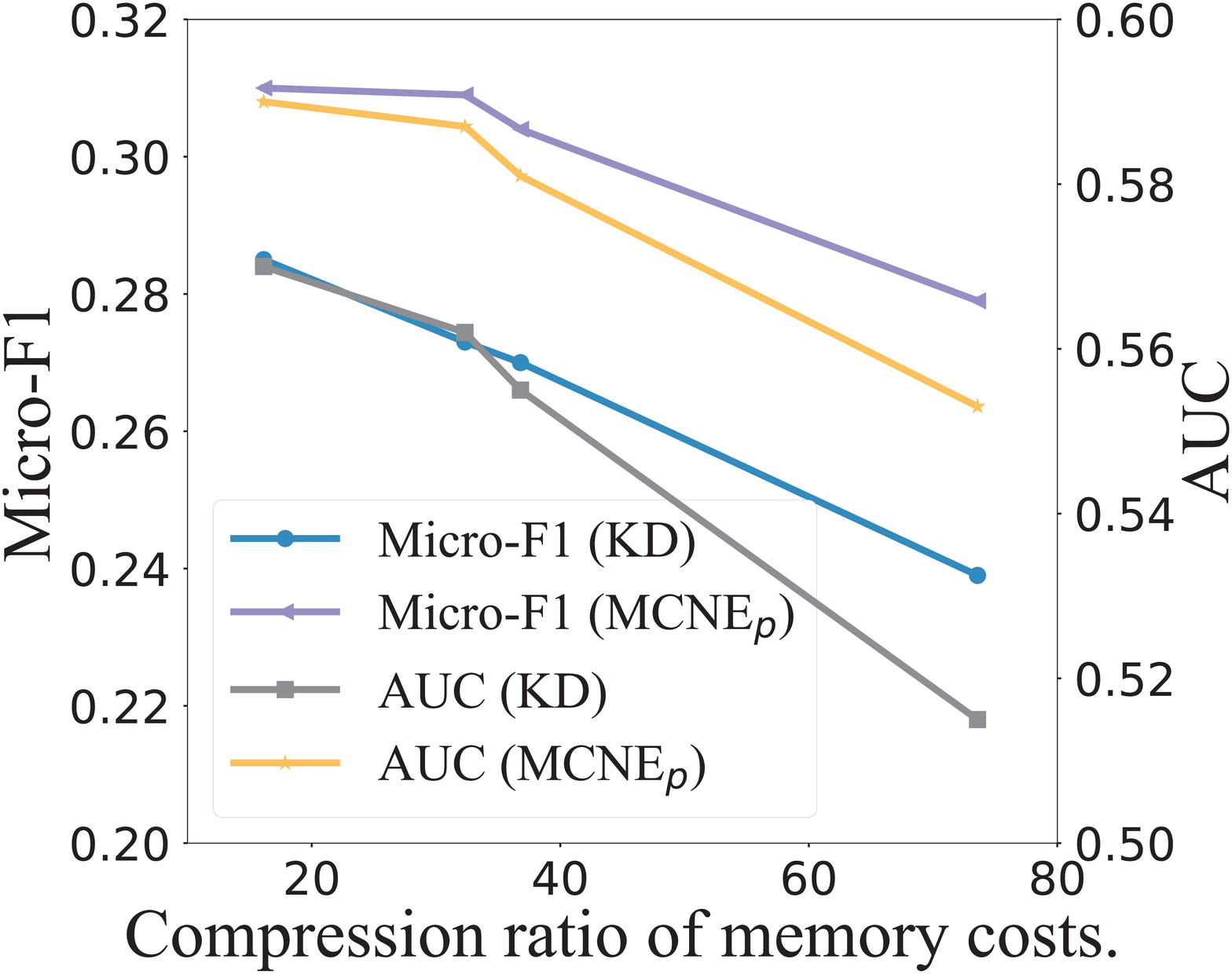}}
	\caption{Performance comparison between KD coding and MCNE$_p$ under different compression ratios of memory cost.}
	\label{fig:compress} 
\end{figure}

As shown in the Figure \ref{fig:framework_kd} and Figure \ref{fig:framework_ml}, the numbers of parameters in the learned embeddings of MCNE$_{p}$ and KD are $(s \times d + |V| \times t)$ and $(K \times D \times d + |V| \times D)$, respectively.
The first part is the size of the shared basis matrix whose entries are float numbers, and the second part is the size of the learned discrete indexes which are integers.
Hence the memory costs of  MCNE$_{p}$ and KD are $(s \times d \times 16 + |V| \times t \times 12)$ bytes and $(K \times D \times d \times 16 + |V| \times D \times 12)$ bytes, respectively.
To ensure the fairness, the number of parameters in KD is same to MCNE$_{p}$: $s = K \times D$ and $t=D$.
From the results, one can obtain the following observations:

\begin{figure}
	\centering
	\includegraphics[width=0.45\textwidth]{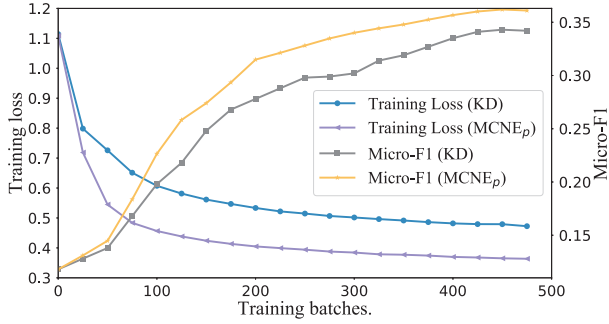}
	\caption{The training trajectories of KD and MCNE$_{p}$.}
	\label{fig:lb} 
\end{figure}

1) Compared with Node2vec, MCNE$_{p}$ achieves comparable performance but the number of parameters is less than 10\% of Node2vec, and also 90\% memories are saved on average. 
By utilizing the partially shared embeddings to reduce the number of continuous basis vectors, MCNE$_{p}$ consumes much less memories. 
In DBLP dataset, MCNE$_{p}$ achieves the highest  compression ratio of parameter number (25.29) and the highest memory compression ratio (32.33). 
MCNE$_{p}$ even achieves higher Micro-F1 score than Node2vec on the DBLP dataset and higher AUC score on the BlogCatalog. This may be because the reconstruction process in the MCNE$_{p}$ can work as the regularization function, which contributes to alleviating the over-fitting issue and thus improves the performance slightly.

2) Compared with the KD coding, MCNE$_{p}$ achieves better performance of both tasks over all the datasets. With the same memory cost, MCNE$_{p}$ outperforms KD by nearly 3\% of the Micro-F1, 2.2\% of the Macro-F1 and 2.5\% of the AUC score.
By removing the blocks, the sharing ratio of the basis vectors is increased, which contributes to improving the representation ability of  MCNE$_{p}$ model and leads to the better performance.

We also compare the performance of KD coding and MCNE$_{p}$ under different compression ratios of memory costs.
We choose different combinations of $s$ and $t$ for MCNE$_{p}$ ($K$ and $D$ for KD) to obtain different compression ratios, and record the corresponding Micro-F1 and AUC scores on the Blogcatalog and DBLP datasets.
Figure \ref{fig:compress} shows the results.
One can see that MCNE$_{p}$ consistently outperforms KD given the same compression ratios.
Besides, the compression ratio of MCNE$_{p}$ is  consistently larger than the ratio of KD coding when both models achieve the same Micro-F1 or AUC score, which proves MCNE$_{p}$ can both reduce memory cost and learn better node embeddings.
One can also see that the performance of both methods decreases with the increase of the compression ratio. 
A larger compression ratio means the model costs fewer memories, but leads to the lower representation capacity of the node embeddings. 
Thus, the tradeoff between the  compression ratio and the quality of node embeddings should be carefully decided.

\noindent \textbf{Learning Process Analysis} 
Next we show the training trajectories of KD and MCNE$_{p}$ by giving the loss curves of both methods on the BlogCatalog dataset as a case study. 
During the model training, we save the parameters after every 25 epochs and finally 20 checkpoint models can be obtained. We also record the training loss and the classification performance (Micro-F1) of each checkpoint model. 
The training loss represents the difference between the learned compact embeddings and the original ones. 
As shown in Figure \ref{fig:lb}, one can see the training loss decreases with the increase of training batches, and a smaller training loss results in a better classification performance. 
One can also see that compared with KD, the loss of MCNE$_{p}$ is consistently smaller leading to a higher Micro-F1 score, which proves MCNE$_{p}$ owns  better feature learning ability.

\begin{figure}
	\centering
	\includegraphics[width=0.40\textwidth]{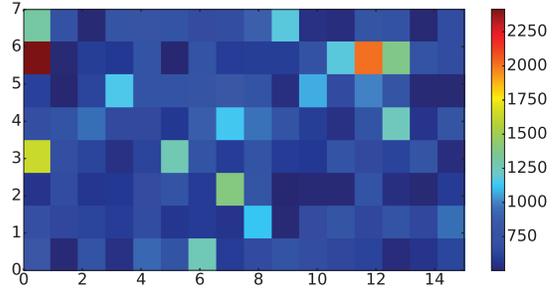}
	\caption{The utilization  efficiency of the basis vectors. Numbers in the x-axis and y-axis are the indexes of basis vectors.}
	\label{fig:codes} 
\end{figure}

\noindent \textbf{Utilization  Efficiency Analysis on Basis Vectors } 
Here we also analyze the utilization efficiency of basis vectors in MCNE$_{p}$ on the BlogCatalog dataset. 
Ideally, all basis vectors should be fully utilized to convey a fraction of latent meaning.
However, as the multi-hot indexes are latently defined and learned, it is possible that some basis vectors may never be used or only be used for very few times, which will cause the resource waste.
We count the occurrence number of each basis vector according to the learned multi-hot indexes, and show the occurrence numbers in Figure \ref{fig:codes}. 
The brightness of the color in a cell is proportion to the occurrence times of the corresponding  basis vector.
One can see that except for several cells with very bright colors, the colors of most other cells are similar, indicating a  balanced distribution of the utilization of the basis vectors. 
It demonstrates that the proposed MCNE$_{p}$ model achieves a desirable utilization efficiency of basis vectors.

\begin{figure}
	\centering
	\includegraphics[width=0.40\textwidth]{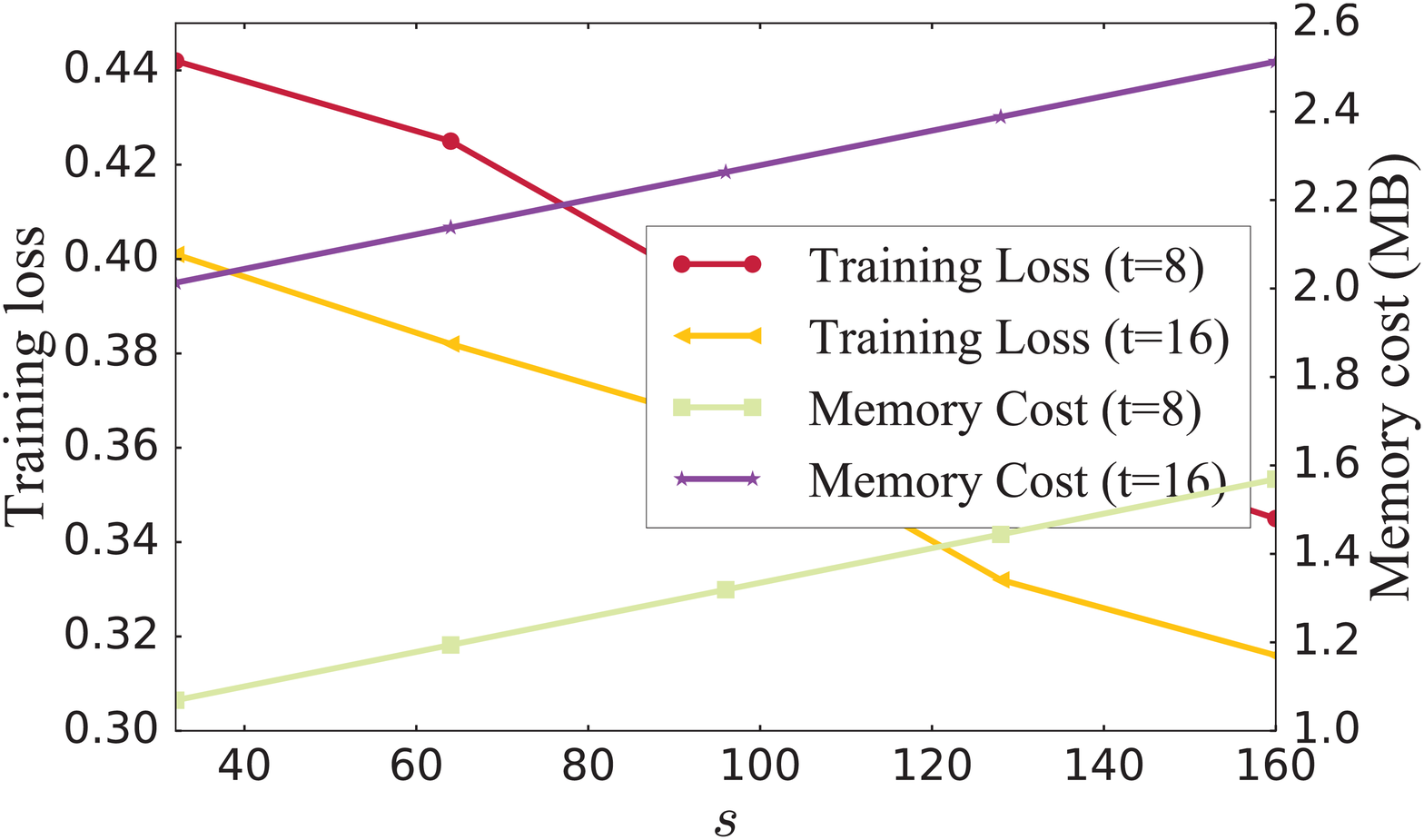}
	\caption{The parameter sensitivity study of MCNE$_{p}$.}
	\label{fig:para1} 
\end{figure}

\begin{table*}
	\centering
	\small
	\begin{threeparttable}
		\caption{Performance comparison among of the end-to-end network embedding models.}
		\begin{tabular}{P{2.0cm}|P{1.2cm}|P{1.2cm}P{1.2cm}P{1.2cm}P{1.2cm}P{2.0cm}P{2.0cm}}
			\toprule
			&Dataset&DeepWalk&LINE&Node2Vec&SDNE&DNE&MCNE$_{t}$ \cr
			\midrule
			Node classi.  &Blog.&0.351&0.347&0.371&0.387&0.207&\textbf{0.397}\cr
			(Micro-F1)&DBLP&0.294&0.295&0.302&0.316&0.146&\textbf{0.342}\cr
			&Flickr&0.324&0.335&0.358&0.369&0.161&\textbf{0.381}\cr
			\midrule
			Node classi.  &Blog.&0.203&0.187&0.224&0.241&0.114&\textbf{0.264}\cr
			(Macro-F1)&DBLP&0.112&0.115&0.126&0.143&0.094&\textbf{0.161}\cr
			&Flickr&0.140&0.151&0.162&0.175&0.102&\textbf{0.182}\cr
			\midrule
			Link Pred. &Blog.&0.732&0.721&0.781&0.773&0.475&\textbf{0.797}\cr
			(AUC) &DBLP&0.566&0.575&0.591&0.617&0.368&\textbf{0.633}\cr
			&Flickr&0.628&0.645&0.683&0.692&0.385&\textbf{0.706}\cr
			\midrule
			\# of para. &Blog.&2.65&2.65&2.65&2.65&2.65 (1.0)&\textbf{0.12 (22.1)}\cr
			(Million) &DBLP&4.30&4.30&4.30&4.30&4.30 (1.0)&\textbf{0.17 (25.3)}\cr
			&Flickr&6.08&6.08&6.08&6.08&6.08 (1.0)&\textbf{0.44 (13.8)}\cr
			\midrule
			\# of  Bytes&Blog.&40.40&40.40&40.40&40.40&\textbf{0.95 (42.5)}&1.44 (28.1)\cr
			(MB) &DBLP&65.63&65.63&65.63&65.63&\textbf{1.54 (42.6)}&2.03 (32.3)\cr
			&Flickr&92.71&92.71&92.71&92.71&\textbf{4.34 (21.4)}&5.33 (17.4)\cr
			\bottomrule
		\end{tabular}
		\label{tab:end2end}
	\end{threeparttable}
\end{table*}

\noindent \textbf{Parameter Sensitivity Analysis} 
We next perform parameter sensitivity study of MCNE$_{p}$ on $s$ and $t$, where $s$ is the number of basis vectors, and $t$ represents how many basis vectors will be selected to form the final node representation. 
Given $t$ is set to 8 and 16, we increase $s$ from 32 to 160, and then record the training losses and the cost memories with different parameter settings.
Figure \ref{fig:para1} shows the results on the BlogCatalog dataset. 
One can see that the training loss decreases with the increase of $s$ or $t$, which means the learned compact embeddings becoming more and more similar to the original ones. However, a larger $s$ means more basis vectors, which will consume more memories and slow down the training speed. A larger $t$ means the final representation is composed by more basis vectors, which costs more calculation resources.

\subsection{Evaluation on End-to-end Compact Network Embedding} 
In this subsection we evaluate the performance of the end-to-end multi-hot embedding model MCNE$_{t}$.
The baselines include the traditional network embedding methods DeepWalk, LINE, Node2Vec and SDNE, and a recent memory-saving model DNE. 
Here we do not compare MCNE$_{t}$ with KD model as KD cannot learn embeddings from the network structure directly.

\noindent \textbf{Quantitative Analysis}
Table \ref{tab:end2end} presents the results of different embedding models along with their parameter numbers and memory costs. 
From the results, we can obtain the following observations:

1) Compared with the traditional embedding methods, MCNE$_{t}$ achieves better performance with much fewer parameters and memory usages on all three datasets. 
Specifically, MCNE$_{t}$ outperforms the best baseline SDNE by 2.5\% on the Micro-F1 classification score with the compression ratio of nearly 20 on average. 
Given a center node, GCN utilizes the weighted aggregates of the features from its neighbors to generate its new representation, which can incorporate the localized topology information to improve the quality of learned embeddings.
By introducing the deep GCN model with stronger feature learning ability, MCNE$_{t}$ outperforms baselines on both datasets.
Overall,  MCNE$_{t}$ can directly learn the multi-hot compact embeddings under the end-to-end scenario, which saves more than 90\% memories while achieves desirable performance.

2) DNE learns the binary codings as the node embeddings. The size of the learned embedding matrix is same to the traditional methods.
Hence, the  count of parameters in the learned embeddings of DNE is $(|V| \times d + |V|)$.
The binary entry can be represented by a bit (0 and 1), while the float number need 16 Bytes (128 bits) in Python. 
Thus DNE has the minimum memory cost. 
However, DNE performs the worst on node classification and link prediction, indicating the learned embedding quality is relatively poor. 
This is reasonable as the binary coding based vectors cannot preserve as much information as the continuous ones.
Different from DNE reduces memory cost by learning binary codings, MCNE$_{t}$ learns partially shared embeddings to reduce the number of continuous vectors, hence  MCNE$_{t}$ has the minimal number of parameters.

\begin{figure*}
	\centering
	\includegraphics[width=0.9\textwidth]{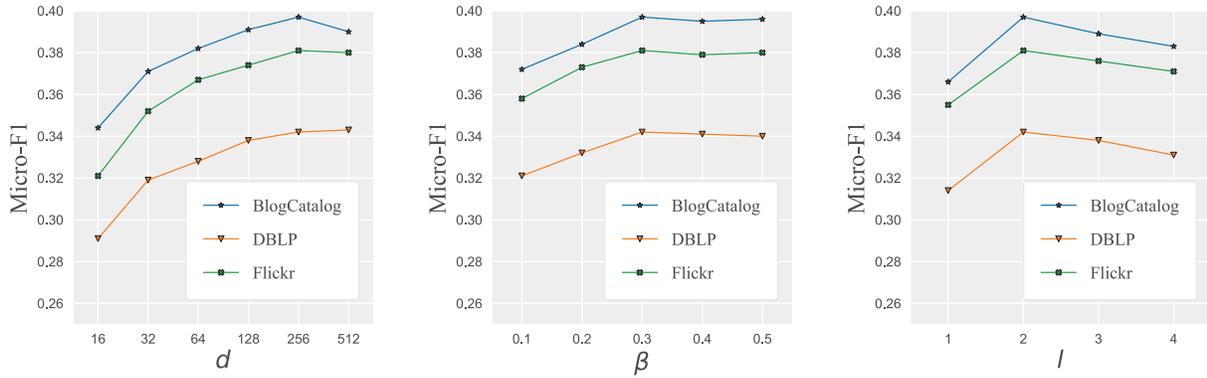}
	\vspace{-1mm}
	\caption{The parameter sensitivity of MCNE$_{t}$.}
	\label{fig:params} 
\end{figure*}

\noindent \textbf{Parameter Sensitivity Analysis}
We next study the performance sensitivities of MCNE$_t$ on the number of GCN layers $l$, the dimension $d$ of the learned embeddings and the reconstruction weight $\beta$.
The performance of  MCNE$_{t}$ on different combinations of $s$ and $t$ is similar to the MCNE$_{p}$ model, which has been discussed in the previous subsection. 
As the Micro-F1, Macro-F1 and AUC scores present similar curves with the change of core parameters, we only report the Micro-F1 score due to space limitation.

The results are shown in Figure \ref{fig:params}.
One can see that, with the increase of the dimension $d$, the Micro-F1 scores first increase and then keep stable. It demonstrates that a larger embedding dimension can provide stronger representation ability when $d$ is comparatively small. 
But when $d$ is too large, the model will encounter the performance bottleneck. 
For the parameter $\beta$, one can see that with the increase of $\beta$, the Micro-F1 scores first increase and then keep stable. It shows that incorporating appropriate reconstruction loss can better preserve the localized topology information from GCN, which contributes to improving the model performance.
Finally for the number of GCN layers $l$, the performance of MCNE$_{t}$ first increases then decreases.
Given a center node, MCNE$_{t}$ model with $l$ GCN layers can aggregate the features of its $l$-hop neighbors to learn the representation.  
An appropriate $l$ value contributes to learning better node representations, but a too large $l$ may introduce noises from long distance neighbors, and thus hurts the performance.

\section{Conclusion}
In this paper, we propose a multi-hot compact network embedding framework to learn the memory-saving node representations. 
We view each node representation as the combination of several basis vectors.
By sharing basis vectors among different nodes, the proposed multi-hot strategy can reduce the number of continuous vectors, which further reduces the memory overhead.
A deep auto-encoder based model MCNE$_{p}$ integrated with a novel component named compressor is proposed to learn compact embeddings from pre-learned node features.
We further propose a graph convolutional network based model MCNE$_{t}$ to  learn compact embeddings in the end-to-end manner.
Experimental results on four real networks demonstrate the significant effectiveness of our proposals.

\section*{Acknowledgments}
This work was supported the Natural Science Foundation of China (Nos. U1636211,61672081,61370126,61602237), Beijing Advanced Innovation Center for Imaging Technology (No.BAICIT-2016001), National Key R\&D Program of China (No.2016QY04W0802 and 2017YFB0802203), Natural Science Foundation of Jiangsu Province (No. BK20171420) and Natural Science Foundation of Guangdong Province (No. 2017A030313334). This work was also supported in part by NSF under grants III-1526499, III-1763325, III-1909323, SaTC-1930941, and CNS-1626432. 